\documentclass[12pt]{article}
\usepackage{amsmath}
\usepackage{amsfonts,epsfig,latexsym}
\usepackage[vcentermath,enableskew]{youngtab}
\usepackage[nosort]{cite}
\def\baselinestretch{1.1}
\newdimen\squaresize \squaresize=12pt
\newdimen\thickness \thickness=0.7pt

\def\square#1{\hbox{\vrule width \thickness
   \vbox to \squaresize{\hrule height \thickness\vss
      \hbox to \squaresize{\hss#1\hss}
   \vss\hrule height\thickness}
\unskip\vrule width \thickness} \kern-\thickness}

\def\cut#1{\hbox{\vrule width-1 \thickness
   \vbox to \squaresize{\hrule height-1 \thickness\vss
      \hbox to \squaresize{\hss#1\hss}
   \vss\hrule height-1\thickness}
\unskip\vrule width +4 \thickness} \kern-\thickness}

\def\vsquare#1{\vbox{\square{$#1$}}\kern-\thickness}

\newcommand{\ft}[2]{{\textstyle\frac{#1}{#2}}}
\def\tilde{\widetilde}
\def\1bar{1\hskip -.275cm -}
\def\2bar{2\hskip -.275cm -}
\def\3bar{3\hskip -.275cm -}

\newsavebox{\uuunit}

\newcommand{\pls}{\!+\!}

\newcommand{\mis}{\!-\!}

\newcommand{\mathon}{\mathversion{bold}}
\newcommand{\mathoff}{\mathversion{normal}}

\newcommand{\ie}{{\it i.e.~}}
\newcommand{\eg}{{\it e.g.~}}
\newcommand{\ap}{\alpha^{\prime}}

\newcommand{\be}{\begin{equation}} \newcommand{\ee}{\end{equation}}
\newcommand{\bea}{\begin{eqnarray}} \newcommand{\eea}{\end{eqnarray}}
\newcommand{\ben}{\begin{displaymath}}
\newcommand{\een}{\end{displaymath}}
 
\newcommand{\nn}{\nonumber} \newcommand{\non}{\nonumber\\}

\makeatletter
\let\old@makecaption=\@makecaption
\def\@makecaption{\small\old@makecaption}
\makeatother

\makeatletter \@addtoreset{equation}{section} \makeatother

\makeatletter
\let\old@startsection=\@startsection
\renewcommand{\@startsection}[6]{\old@startsection{#1}{#2}{#3}{#4}{#5}{#6\mathversion{bold}}}
\makeatother

\let\oldPhi=\Phi
\let\oldPsi=\Psi
\let\oldGamma=\Gamma
\let\oldSigma=\Sigma
\renewcommand{\Phi}{\mathnormal{\oldPhi}}
\renewcommand{\Psi}{\mathnormal{\oldPsi}}
\renewcommand{\Gamma}{\mathnormal{\oldGamma}}
\renewcommand{\Sigma}{\mathnormal{\oldSigma}}

\newcommand{\hypref}[2]{\ifx\href\asklfhas #2\else\href{#1}{#2}\fi}

\newcommand{\mult}{\mathcal{V}}

\newcommand{\superN}{\mathcal{N}}
\newcommand{\gym}{g_{\scriptscriptstyle\mathrm{YM}}}

\newcommand{\bigbrk}[1]{\bigl(#1\bigr)}

\newcommand{\nl}{\nonumber\\&&\mathord{}}

\newcommand{\earel}[1]{\mathrel{}&#1&\mathrel{}}
\newcommand{\eq}{\earel{=}}
\newenvironment{myeqnarray}{\arraycolsep0pt\begin{eqnarray}}{\end{eqnarray}\ignorespacesafterend}
\newenvironment{myeqnarray*}{\arraycolsep0pt\begin{eqnarray*}}{\end{eqnarray*}\ignorespacesafterend}

\def\[{\begin{equation}}
\def\]{\end{equation}}
\def\<{\begin{myeqnarray}}
\def\>{\end{myeqnarray}}

\begin{document}

\thispagestyle{empty}

\begin{center}

{\small\ttfamily ROM2F/04/26\hspace*{0.8cm} }

\end{center}


\begin{center}

\renewcommand{\thefootnote}{\fnsymbol{footnote}}

{\mathon\bf\Large  Higher spin symmetry (breaking) \\ in ${\cal
N}=4$ SYM and holography\footnote{Talk delivered at ``Strings 04''
in Paris, 28 june 2004.}
\par \mathoff}%
\bigskip\bigskip

\addtocounter{footnote}{1} \textbf{
M.~Bianchi\footnote{\texttt{Massimo.Bianchi@roma2.infn.it}}}

\textit{
Dipartimento di Fisica and INFN \\
Universit\`a di Roma ``Tor Vergata''\\
00133 Rome, Italy}\par

\setcounter{footnote}{0}
\end{center}

\begin{abstract}
I concisely review the results of
\cite{Bianchi:2003wx,Beisert:2003te,Beisert:2004di} that shed some
light on {\it ``La Grande Bouffe"}\footnote{Several people asked
me the origin of this terminology. It is the title of a movie
directed by Marco Ferreri, interpreted, among others, by Marcello
Mastroianni and Ugo Tognazzi and presented in 1973 at {\it
Festival du Cinema} in Cannes where it received the International
Critics Award.}, the pantagruelic Higgs mechanism, whereby HS
gauge fields in the AdS bulk eat lower spin Goldstone fields, and on its 
holographic implications such as the emergence of anomalous dimensions in the 
boundary N=4 SYM theory.  
\end{abstract}

\def\baselinestretch{1.1}

 Formulating the dynamics of higher spin
(HS) fields is a long standing problem, see \eg
\cite{Vasiliev:2004qz, Sorokin:2004ie, Bouatta:2004kk} and
references therein\footnote{For lack of space, I will often need
to refer to comprehensive review papers rather than the original
literature. I apologize for this inconvenience and warmly invite
the interest reader to consult the exhaustive list of reference in
\cite{BianchiRTN}.}. In the massless bosonic case, Fronsdal wrote
down linearized field equations for totally symmetric tensors
$\varphi^{(\mu_1...\mu_s)}$ that in $D=4$ arise from a Lorentz
covariant (quadratic) action upon imposing `double tracelessness'
$\eta^{\mu_1\mu_2} \eta^{\mu_3\mu_4}\varphi_{\mu_1...\mu_s}=0$. HS
gauge invariance corresponds to restricted transformations $ \delta
\varphi_{\mu_1...\mu_s} = \partial_{(\mu_1}
\epsilon_{\mu_2...\mu_s)} $ with traceless parameters. Fang and
Fronsdal then extended the analysis to fermions, while Singh and
Hagen formulated equations for massive fields that reduce to
Frondal's or Fang-Fronsdal's in the massless limit upon removing
certain auxiliary fields. String theory in flat spacetime can be
considered as a theory of an infinite number of HS gauge fields of
various rank and (mixed) symmetry in a broken phase. At high
energies these symmetries should be restored resulting in a new
largely unexplored phase. Upon coupling HS fields to (external)
gravity, the presence of the Weyl tensor in the variation of the
action for $s>2$, resulting from the Riemann tensor in the
commutator of two covariant derivatives, spoils HS gauge
invariance even at the linearized level and for on-shell
gravitational backgrounds, except for spin $s\le 2$, where at most
the Ricci tensor appears. Problems with interactions for HS gauge
fields in flat spacetime are to be expected since the
Coleman - Mandula theorem and its generalization by Haag -
Lopusanski - Sohnius imply a trivial S-matrix whenever the
Poincar\'e group is extended by additional spacetime generators
such as HS symmetry currents. Moreover closure of the HS algebra
requires an infinite tower of symmetries as soon as HS fields with
$s>2$ enter the game. A completely new approach to the
interactions, if any, is to be expected in order to deal with an
infinite number of HS fields and arbitrarily high derivatives
\cite{Vasiliev:2004qz, Sorokin:2004ie, Bouatta:2004kk}.

According to Fradkin and Vasiliev, the situation improves
significantly when the starting point is taken to be maximally
symmetric AdS space\footnote{Results for dS space can be formally
obtained by analytic continuation.} with non-vanishing
cosmological constant $\Lambda=-(D-2)(D-1)/R^2$ rather than flat
spacetime. One can then use the HS analogue of MacDowell -
Mansouri - Stelle - West (MDMSW) $SO(D-1,2)$ formulation of gravity
in order to keep HS gauge symmetry manifest and compactly organize
the resulting higher derivative interactions and the associated
non-locality. Vasiliev has been able to pursue this program till
the very end, \ie at the fully non-linear level, for massless
bosons in $D=4$ \cite{Vasiliev:2004qz}. In Vasiliev's equations
$\Lambda$ plays a double role. On the one hand it organizes higher
derivative interactions, very much like the string scale $M_s =
1/\sqrt{\ap}$ in string theory. On the other hand it allows one to
define a generalized $SO(D-1,2)$ curvature that vanishes exactly
for AdS. The AdS/CFT correspondence at the HS enhancement point
seems exactly what the  doctor ordered. At generic radius $R$,
superstring theory should describe HS fields in a broken phase. At
some critical radius, Vasiliev's equations or some generalization
thereof should govern the dynamics of the exactly massless phase.

In the much studied case of ${\cal N}=4$ SYM theory in $d=4$ with
$SU(N)$ gauge group, the holographic correspondence with Type IIB
superstring theory on $AdS_5 \times S^5$ with $N$ units of RR
5-form flux has been an unprecedented source of insights
\cite{Aharony:1999ti, D'Hoker:2002aw, Bianchi:2000vh,
Tseytlin:2003ii}. At vanishing coupling, ${\cal N}=4$ SYM exposes
HS Symmetry enhancement \cite{WittenJHS60, Sundborg:1999ue,
Polyakov:2001af}. Conformal invariance indeed implies that a spin
$s$ current, such as $ J_{({\mu_1}{\mu_2}... {\mu_s})|} = Tr
(\varphi_i D_{(\mu_1}... D_{\mu_{s})|}\varphi^i) +...$ saturating
the unitary bound $\Delta_0 = 2 + s$ be conserved. ${\cal N}=4$
superconformal symmetry $(P)SU(2,2|4)$ implies that twist two
operators are either conserved currents or superpartners thereof
\cite{DobrevPetkova, Dolan:2002zh, Heslop:2003xu}. Altogether they
form the {\it doubleton} representation of $HS(2,2|4)$, the HS
extension of $(P)SU(2,2|4)$. The weak coupling regime on the
boundary should be holographically dual to a highly stringy regime
in the bulk, where the curvature radius $R$ is small in string
units $R \approx \sqrt{\ap}$ and the string is nearly tensionless.
Although quantizing the superstring in $AdS_5\times S^5$ is a
difficult and not yet accomplished task \cite{Berkovits:2002zk},
Sezgin and Sundell have been able to write down linearized field
equations for the `massless' $HS(2,2|4)$ doubleton
\cite{Sezgin:2001zs, Sezgin:2001yf, Sezgin:2002rt}. As in
Vasiliev's case the field content can be assembled into a master
connection $A$ and a master scalar (curvature) $\Phi$. The former
transform in the adjoint representation of $HS(2,2|4)$ and
contains physical gauge fields with $s\ge 1$ and charge $B=0,\pm
1$. The latter transform in the twisted adjoint representation and
contributes physical fields with spin $s\le 1/2$ or $s\ge 1$ and
charge $|B|\ge 3/2$  such as self-dual two-form potentials. The
field strengths $ F_A = dA + A\wedge *A$ and $D_A \Phi = d\Phi + A
* \Phi -\Phi*\tilde{A} $ transform covariantly  $ \delta F_A =
[F_A,\epsilon]_*$, $\delta D_A\Phi = D_A\Phi * \tilde\epsilon -
\epsilon *D_A\Phi$ under HS gauge transformations $\delta A = d
\epsilon + [A,\epsilon]_*$, whereby $\delta \Phi = \Phi *
\tilde\epsilon - \epsilon *\Phi$. The linearized constraints and
integrability conditions lead after some tedious algebra to the
correct linearized field equations for the `matter' fields with
$s\le 1/2$, for the HS gauge fields and for the antisymmetric
tensors with generalized self-duality \cite{Sezgin:2001zs,
Sezgin:2001yf, Sezgin:2002rt}.

Possibly because of the presence of these generalized self dual
tensors, an interacting $HS(2,2|4)$ gauge theory has not yet been
formulated. In some sense, however, non-linear equations of
Vasiliev's type encode combinatorial interactions which are
present even in a free field theory, where HS symmetry is
unbroken, or couplings to multi-particle states at finite
$N$\footnote{Precisely for this reason they are relevant in the
$d=3$ $O(N)$ model on the boundary of $AdS_4$ \cite{PetkouRTN}.}.
Although truncation to the HS massless multiplet (doubleton)
should be consistent at the point of HS enhancement this should no
more be the case for generic $R$. When interactions are turned on,
\ie at $\lambda \neq 0$, only a handful of HS fields remain
massless. The vast majority partecipates in a pantagruelic Higgs
mechanism, termed {\it ``La Grande Bouffe"} in
\cite{Bianchi:2003wx,Beisert:2003te,Beisert:2004di}, whereby HS
gauge fields eat lower spin Goldstone fields. In the dual ${\cal
N}=4$ SYM description only the 1/2 BPS short multiplets, 
corresponding to ${\cal N}=8$ gauged
supergravity and its Kaluza-Klein (KK) recurrences, are protected
against quantum corrections to their dimensions. Except for the
$\superN=4$ supercurrent multiplet, the infinite tower of
conserved doubleton multiplets acquire anomalous dimensions which
violate the conservation of the HS currents at the quantum level.
At one-loop, one has \cite{Beisert:2004ry}
\[
\gamma_{\rm 1-loop}(2n) =\frac{\gym^2 N}{2\pi^2}\,h(2n),\qquad
h(j)=\sum_{k=1}^j\frac{1}{k},
\]
This elegant `number theoretic' formula gives a clue on how to
compute generic anomalous dimensions at first order in
perturbation theory relying on HS symmetry breaking considerations
and leads to integrability of the super spin chain Hamiltonian
that represents the action of the (one-loop) dilatation operator
\cite{Beisert:2004ry}. For the purpose of describing the breaking
of $HS(2,2|4)$ to $PSU(2,2|4)$ one should identify the Goldstone
multiplets that provide the `longitudinal' lower spin modes to the
massless HS doubleton and hopefully determine their couplings by
imposing (linearized) HS symmetry. At large $N$, this problem
might turn out to be easier to solve than constructing a fully
non-linear massless $HS(2,2|4)$ theory because it should only
require a little bit more than the knowledge of the linearized
field equations. For the (long) ${\cal N}=4$ Konishi multiplet
\cite{Bianchi:2001cm, Andrianopoli:1998ut}, with $2^{16}$
components, dual to the first massive level of string excitations,
for instance one expects a decomposition like \be {\cal K}_{long}
\leftrightarrow {\cal K}_{short} + {\cal K}_{1/4} + {\cal K}_{1/8}
+ {\cal K}^*_{1/8} \ee \ie the semishort multiplet ${\cal
K}_{short}$, belonging to the HS `doubleton', eats the lower spin 
Goldstone multiplets ${\cal
K}_{1/8}$, ${\cal K}^*_{1/8}$ and ${\cal K}_{1/4}$, belonging to
the `massive' HS multiplets associated to the totally
antisymmetric `tripleton' and the window-like `tetrapleton'.

Group theoretic analysis and the pp-wave limit allow us not only to determine
the HS content of the free ${\cal N}=4$ SYM spectrum at large $N$
\cite{Beisert:2004di} but also to match it with the superstring
spectrum extrapolated to the point of HS symmetry enhancement
\cite{Bianchi:2003wx, Beisert:2003te}. Even at finite coupling,
nearly BPS operators with large R-charge $J$ and dimension $\Delta
\approx J$ can be built by successively inserting {\it impurities}
inside Chiral Primary Operators (CPO's) of the form $Tr(Z^J)$.
Berenstein, Maldacena and Nastase (BMN) argued that the sector
with $J\approx \sqrt{N}$ is described by type IIB superstring on
the maximally supersymmetric pp-wave emerging from a Penrose limit
of $AdS_5\times S^5$ \cite{Berenstein:2002jq}. Despite the
presence of a null RR 5-form flux $F_{+1234} = F_{+5678}=\mu$,
superstring fluctuations can be quantized in the light-cone gauge,
where $p^+ = J/\mu \ap$. The spectrum of the light-cone
Hamiltonian \be H_{LC} = p^- = \mu (\Delta - J) = \mu \sum_n N_n
\omega_n \quad , \quad \omega_n = \sqrt{ 1 + 
\frac{n^2 \lambda}{J^2}} \quad , 
\label{bmnformula} \ee with the level matching
condition $\sum_n n N_n=0$, represents a prediction for the
spectrum of anomalous dimensions of the so-called BMN operators,
that form $(P)SU(2,2|4)$ multiplets at large but finite $J$. For
our purposes it is crucial that any single-trace operator in
${\cal N}=4$ SYM be identified with some BMN operator with an
arbitrary but finite number of impurities.

In flat spacetime, the single-particle type IIB superstring
spectrum results from  combining left- and right-moving modes with
the same chirality projection on the vacuum and imposing
level-matching $\ell = \sum n N^L_n = \sum_n n N^R_n$. In the
light-cone, where only $SO(8)\subset SO(9,1)$ is manifest, the
chiral groundstates $|{\cal Q}\rangle_{L/R}$ consists of ${\bf
8}_V$ bosons and ${\bf 8}_S$ fermions. At $\ell = 0$ one finds the
`transverse' modes of type IIB $N=(2,0)$ supergravity $({\bf
8}_V-{\bf 8}_S)\times ({\bf 8}_V-{\bf 8}_S)$. At higher levels,
$\ell\geq 1$, the (chiral) spectrum assembles into full
representations of the massive transverse Lorentz group $SO(9)$.
For instance at $\ell = 1$,  one finds ${\bf 44} + {\bf 84} - {\bf
128}$ of $SO(9)$, corresponding to a symmetric tensor (`spin 2'),
a 3-index totally antisymmetric tensor and a spin-vector (`spin
3/2'). At higher levels the situation is similar. Actually the spectrum
can be organized into ${\cal N}=(2,0)$ supermultiplets,
whose groundstates are annihilated by half of the $32$
supercharges. For $\ell=1$ the groundstate cannot be other than an
$SO(9)$ singlet $V^{L/R}_{\ell=1} = {\bf 1}$, \ie a scalar, since
$2^8\times 2^8 = 2^{16}$ equals the number of d.o.f. at this
level. At higher levels the situation is not so straightforward,
but one can eventually deduce a recurrence relation that yields $
V^{L/R}_{\ell=1}= {\bf 1} , V^{L/R}_{\ell=2} = {\bf 9} ,
V^{L/R}_{\ell=3} = {\bf 44}- {\bf 16}, ... $ for the first few
levels. In summary, the Hilbert space of type IIB superstring
excitations in flat space can be written as \be {\cal H}_{flat} =
{\cal H}_{sugra} + {\cal T}_{susy} \sum_\ell V^{L}_{\ell} \times
V^{R}_{\ell} \ee where ${\cal T}_{susy}$ represents the action of
the 16 `raising' supercharges. States with maximum spin
$s_{Max}=2\ell + 2$ at level $\ell$ belong to the first Regge
trajectory which is generated by oscillators with lowest
non-trivial mode number. Moreover, the partial sums
$\sum_\ell^{1,K} V^{L}_{\ell} \times V^{R}_{\ell}$ form $SO(10)$
multiplets. This is related to the possibility of `covariantizing'
the massive spectrum of type IIB, which is identical to the one of
type IIA, to $SO(10)$, by lifting it to $D=11$ \cite{Bars:2004dg},
or to $SO(9,1)$, by introducing worldsheet (super)ghosts
\cite{Berkovits:2002zk}.

In order to extrapolate the massive string spectrum from flat
space to $AdS_5\times S^5$ at the HS symmetry enhancement point
one should first decompose $SO(9)$ into $SO(4)\times SO(5)$, the
relevant stability group of a massive particle. This
straightforwardly determines two of the quantum numbers of the
$(P)SU(2,2|4)$ superisometry group, namely the two spins $(j_L,
j_R)$ of $SO(4)\subset SO(4,2)$. The set of allowed
representations of the $S^5$ isometry group $SO(6)\approx SU(4)$
are the ones that contain a given representation of $SO(5)$ under
the decomposition $SO(6) \rightarrow SO(5)$. Denoting irreps by
their Dynkin labels, $[m,n]$ for $SO(5)$ and $[k,p,q]$ for
$SO(6)$, group theory yields the KK towers \be {\rm KK}_{[m,n]} =
\sum_{r=0}^{m}\sum_{s=0}^{n} \sum_{p=m-r}^\infty \left[r\pls
s,\,p,\,r\pls n\mis s\right] +\sum_{r=0}^{m-1} \sum_{s=0}^{n-1}
\sum_{p=m-r-1}^\infty \left[r\pls s\pls1 ,\,p,\,r\pls n\mis
s\right] \;. \label{KK56} \ee Any ambiguity in the lift, say, of
the (pseudo-real) spinor ${\bf 4}$ of $SO(5)\approx Sp(4)$ to the
complex ${\bf 4}$ of $SO(6)\approx SU(4)$ or to its complex
conjugate ${\bf 4}^*$ is resolved by the infinite sum over KK
recurrences. Once the $SO(4)\times SO(6)$
 quantum numbers are determined, the perturbative
superstring spectrum turns out to be encoded in \be {\cal H}_{AdS}
= {\cal H}_{sugra} + {\cal T}_{KK} {\cal T}_{susy} \sum_\ell
V^{L}_{\ell} \times V^{R}_{\ell} \label{adsspect}\ee where ${\cal
T}_{KK} = \sum_p [0,p,0]$ represents KK towers that boil down to
sums over scalar spherical harmonics, \ie $p$-fold symmetric
traceless tensors of $SO(6)$. ${\cal T}_{susy}$ represents the
action of the 16 `raising' Poincar\'e supercharges $Q$ and
$\bar{Q}$. $V^{L/R}_{\ell}$, defined in flat space, are to be
decomposed under $SO(4)\times SO(5)$ and lifted to $SO(4)\times
SO(6)$. Formula (\ref{adsspect}) looks deceivingly simple, almost
trivial, since the most interesting information, the scaling
dimension $\Delta_0$ at the HS enhancement
point is still missing.

So far we have tacitly assumed that there are no non-perturbative
states that can appear in the single-particle spectrum as a result
of strings or branes wrapping non trivial cycles. Indeed there are
no such states with finite mass at small $g_s$,\ie large $N$,
since the only non trivial cycles of $S^5$ are a 0-cycle (a point)
or a 5-cycle (the full space). Although there can be ambiguities
in extrapolating the perturbative spectrum from large radius,
where KK technology is reliable, to small radius, where HS
symmetry is restored, we expect `level repulsion' at large $N$
\cite{Polyakov:2001af}. This guarantees that any state identified
at large radius (strong 't Hooft coupling) can be smoothly if not
explicitly followed to small radius (weak coupling) since
trajectories of different fields/operators with the same quantum
numbers never intersects.

One can thus start with identifying the string excitations that
are expected to become massless at the point of enhanced HS
symmetry \cite{Sezgin:2001zs, Sezgin:2001yf, Sezgin:2002rt}. In
particular the totally symmetric and traceless tensors of rank
$2\ell - 2$ at level $\ell>1$ appearing in the product of the
groundstates $V^{L}_{\ell} \times V^{R}_{\ell}$ become massless
and thus correspond to the sought for conserved HS currents on the
boundary if one assigns them $\Delta_0 = 2\ell$, that works fine
for $\ell=1$, too. The states with $PSU(2,2|4)$ quantumm numbers
$\{2\ell;(\ell-1,\ell-1);[0,0,0]\}$ are HWS's of semishort
multiplets \cite{DobrevPetkova, Dolan:2002zh, Heslop:2003xu,
Andrianopoli:1998ut}. Moreover the KK recurrences of these states
at floor $p$ arising from the action of ${\cal T}_{KK}$ are
naturally assigned \cite{Bianchi:2003wx} \be \Delta_0 = 2\ell + p
\label{simple}\ee which represents the $PSU(2,2|4)$ unitary bound
for a spin $s= 2\ell - 2$ current in the $SO(6)$ irrep with Dynkin
labels $[0,p,0]$. It is remarkable how simply assuming HS symmetry
enhancement fixes the AdS masses, \ie scaling dimensions, of a
significant fraction of the spectrum. Nevertheless even at this
particularly symmetric point there should be operators / states
well above the unitary bounds. Surprisingly, (\ref{simple}) turns
out to be correct for all primary states with mass/dimension
$\Delta_0 \le 4$. Notice that `commensurability' of the two
contributions -- spin $s\approx \ell$ and KK `angular momentum' $J
\approx p$ -- suggests that $R = \sqrt{\ap}$, for what this might
mean. In order to find a mass formula that could extend and
generalize (\ref{simple}), it is convenient to take the
 BMN formula (\ref{bmnformula}) as a hint. Although derived under
the assumptions of large $\lambda$ and $J$
\cite{Berenstein:2002jq}, there seems to be no serious problem in
extrapolating it to finite $J$ at vanishing $\lambda$, where
$\omega_n = 1$ for all $n$. Indeed, Niklas Beisert has shown that
(two-impurity) BMN operators form $PSU(2,2|4)$ multiplets at
finite $J$ and are thus amenable to the extrapolation
\cite{Beisert:2004ry}. The resulting formula reads
\cite{Beisert:2003te} \be \Delta_0 = J + \nu \label{magic}\ee where
$\nu = \sum_n N_n$ is the number of oscillators applied to the
`vacumm' $|J=\mu\ap p^+\rangle$ and $J$ is the $U(1)$ charge in
the decomposition of $SO(10)$ into $SO(8)\times U(1)_J$, where
$SO(8)$ is the massless little group. In turn, $SO(10)$ arises
from `covariantizing' $SO(9)$. Although cumbersome, the procedure
is straightforward and can be easily implemented on a computer.
Given the $SO(10)$ content of the flat space string spectrum,
equation~\eqref{magic} uniquely determines the dimensions
$\Delta_0$ of the superstring excitations around $AdS_5\times S^5$
at the HS point. The case $\ell=1$ is almost trivial, as an
illustration, let us thus consider the string levels $\ell=2,3$:
\bea V_2 \eq [1,0,0,0,0]^2-[0,0,0,0,0]^3 \non
\earel{\stackrel{SO(8)\times SO(2)}{\rightarrow}}
 [1,0,0,0]^2_0+
[0,0,0,0]^2_1+[0,0,0,0]^2_{-1} -[0,0,0,0]^3_0\label{second}\\
\earel{\stackrel{\eqref{magic}}{\rightarrow}}
 [1,0,0,0]_2+ [0,0,0,0]_1 \;,
\eea \bea  V_3 \eq [2,0,0,0,0]^3-[1,0,0,0,0]^4-[0,0,0,0,1]^{5/2}
\nonumber \\ \earel{\stackrel{SO(8)\times SO(2)}{\rightarrow}}
  [2,0,0,0]^3_0+
[1,0,0,0]^3_1+[1,0,0,0]^3_{-1} +[0,0,0,0]^3_0+ [0,0,0,0]^3_2 \nl
+[0,0,0,0]^3_{-2}
 -  [1,0,0,0]^4_0- [0,0,0,0]^4_1-[0,0,0,0]^4_{-1} \nl
-[0,0,0,1]^{5/2}_{1/2}-[0,0,1,0]^{5/2}_{-1/2} \nonumber \\
\earel{\stackrel{\eqref{magic}}{\rightarrow}}
 [2,0,0,0]_3 +[1,0,0,0]_2
+[0,0,0,0]_1 -[0,0,1,0]_{3}-[0,0,0,1]_{2} \;. \eea

With the above assignments of $\Delta_0$, negative multiplicities
are harmless since they cancel in the sum over KK recurrences
after decomposing $SO(10)$ w.r.t. $SO(4)\times SO(6)$. For these
low massive levels, the conformal dimensions determined
by~\eqref{magic} all saturate $SO(10)$ unitary bounds of the form
$\Delta_\pm= 1+ k + 2l + 3m + 2(p+q)\pm(p-q)/2$. At higher levels,
starting from a $\Delta_0=3$ singlet at level $\ell=5$, this
bound is still satisfied but no longer saturated: the correct
conformal dimensions are rather obtained from \eqref{magic}. The
results for string levels $\ell=4$ and $\ell=5$ are displayed in
the following tables and organized under $SO(10)\times
SO(2)_{\Delta_0}$, with Dynkin labels $[k,l,m,p,q]_{\Delta_0}$ and
$[k,l,m,p,q]^* \equiv [k,l,m,p,q]-[k\!-\!1,l,m,p,q]$.

          \noindent$\ell=4$ :
     {\footnotesize\begin{equation}\nonumber\begin{tabular}{l|l}
     $\Delta_0$ & ${\cal R}$ \\ \hline
     $4$ &   $
     [3, 0, 0, 0, 0]^*
     $\\ \hline
     $\frac{7}{2}$ &   $
     [1, 0, 0, 0, 1]^*
     $\\ \hline
     $3$ &   $
     [0, 1, 0, 0, 0]
     $\\ \hline
     \end{tabular}\end{equation}}

     \noindent$\ell=5$ :
     {\footnotesize\begin{equation}\nonumber\begin{tabular}{l|l}
     $\Delta_0$ & ${\cal R}$ \\ \hline
     $5$ &   $
     [4, 0, 0, 0, 0]^*
     $\\ \hline
     $\frac{9}{2}$ &   $
     [2, 0, 0, 0, 1]^*
     $\\ \hline
     $4$ &   $
     [0, 0, 1, 0, 0]
     +[1, 1, 0, 0, 0]^*
     $\\ \hline
     $\frac{7}{2}$ &   $
     [1, 0, 0, 0, 1]
     $\\ \hline
     $3$ &   $
     [0, 0, 0, 0, 0]
     $\\ \hline
     \end{tabular}\end{equation}}

In order to test the above prediction for the single-particle
superstring spectrum on $AdS_5\times S^5$ at the HS point against
the spectrum of free ${\cal N}=4$ SYM theory at large $N$, one has
to devise an efficient way of computing gauge-invariant single
trace operators \cite{Sundborg:1999ue, Polyakov:2001af,
Bianchi:2003wx}. For an $SU(N)$ gauge group this means taking care
of the ciclicity of the trace in order to avoid multiple counting.
Moreover one should discard operators which would vanish because
of the field equations and deal with the statistics of the
elementary fields. The mathematical tool one has to resort to is
Polya theory that allows one to count `words' $A, B, ...$ of a
given `length' $n$ composed of `letters' chosen from a given
`alphabet' $\{a_i\}$, modulo some symmetry operation: $A\approx B$
if $A=g B$ for $g\in {\cal G}$. In order to compute Polya cycle
index it is convenient to decompose the discrete group ${\cal
G}\subset {\cal S}_n$ into conjugacy classes whose representatives
$[g]=(1)^{b_1(g)} (1)^{b_1(g)}...(n)^{b_n(g)}$ are characterized
by the numbers $b_k(g)$ of cycles of length $k$. For cyclic
groups, ${\cal G}=Z_n$, conjugacy classes are labelled by divisors
$d$ of $n$, $[g]_d=(d)^{n/d}$, and the cycle index simply reads
\be {\cal P}_{_{Z_n}}(\{a_i\}) = \frac{1}{n}\sum_{d|n} {\cal
E}(d)\left(\sum_i a_i^d\right)^{n/d} \label{polyah}\ee where
${\cal E}(d)$ is Euler's totient function which counts the number
of elements in  the conjugacy class $[g]_d$. ${\cal E}(d)$ equals
the number of integers relatively prime to and smaller than $d$,
with the understanding that ${\cal E}(1)=1$, and satisfies
$\sum_{d|n} {\cal E}(d)=n$.

For ${\cal N}=4$ SYM the alphabet is given by the elementary
fields and their derivatives (modulo the field equations)
$\{\partial^k \varphi,
\partial^k \lambda,
\partial^k F\}$ that transform in the
{\it singleton} representation of $PSU(2,2|4)$. As a first step,
one computes the on-shell single letter partition function or
rather the Witten index ${\cal Z}_1(q) = Tr(-)^F q^\Delta_0$, in
order to take statistics into account. For a single (abelian)
${\cal N}=4$ vector multiplet, one has \be {\cal Z}_1(q)= 2 q{(3
+\sqrt{q})\over (1+\sqrt{q})^3} \quad . \label{singpf} \ee
Plugging (\ref{singpf}) into (\ref{polyah}) one finds the
single-trace partition function \cite{Sundborg:1999ue,
Polyakov:2001af, Bianchi:2003wx}
 \bea {\cal Z}_{{\cal N}=4}(q) \eq
 \sum_{n=2}^\infty \sum_{n|d}
\frac{{\cal E}(d)}{n}\,\left[
\frac{2q(3+q^{\frac{d}{2}})}{(1+q^{\frac{d}{2}})^3}\right]^{\frac{n}{d}}
\label{polyahA}\\
\eq 21\,q^2 - 96\,q^{\frac{5}{2}} + 376\,q^3 -
1344\,q^{\frac{7}{2}} + 4605\,q^4 - 15456\,q^{\frac{9}{2}} +
  52152\,q^5
\nl - 177600\,q^{\frac{11}{2}} + 608365\,q^6 -
2095584\,q^{\frac{13}{2}} + 7262256\,q^7 -
25299744\,q^{\frac{15}{2}} \nl + 88521741\,q^8 -
310927104\,q^{\frac{17}{2}} + 1095923200\,q^9 -
3874803840\,q^{\frac{19}{2}} \nl + 13737944493\,q^{10} +
\mathcal{O}(q^{\frac{21}{2}}) \; \eea  for $SU(N)$ at large $N$,
where mixing with multi-trace operators is suppressed.

In order to identify superconformal primaries, one can pass ${\cal
Z}_{{\cal N}=4}(q)$ through an Eratosthenes super-sieve, that
removes superdescendants. This task can be accomplished by first
subtracting 1/2 BPS multiplets
\be {\cal Z}_{BPS}(q) = \frac{q^2\bigbrk{20+80q^{\frac{1}{2}}
+146q+144q^{\frac{3}{2}}+81q^2+24q^{\frac{5}{2}}+3q^3}}{(1-q)
(1+q^{\frac{1}{2}})^8} \;, \ee
from (\ref{polyahA}) and then dividing by \be {\cal T}_{SO(10,2)}(q) =
(1-q^2)\frac{(1-q^\frac{1}{2})^{16}}{
 (1-q)^{10}} = {\cal T}_{susy}(q) {\cal T}_{KK}(q).
 \ee
 that not only removes superdescendants generated by
 ${\cal T}_{susy} = {(1-q^\frac{1}{2})^{16}}/{
 (1-q)^{4}}$
 but also the operators dual to the KK recurrences generated by
${\cal T}_{KK}= {(1-q^2)}/{
 (1-q)^{6}}$,
 where the numerator implements the $SO(6)$ tracelessness condition.
 For ${\cal Z}_{SO(10,2)}(q)= [
  {\cal Z}_{{\cal N}=4}(q)-{\cal Z}_{BPS}(q)]/
{\cal T}_{SO(10,2)}(q)$
one eventually finds the expansion  \bea
{\cal Z}_{SO(10,2)}(q)\eq
 q^2 + 100\,q^4 + 236\,q^5 - 1728\,q^{\frac{11}{2}} + 4943\,q^6
- 12928\,q^{\frac{13}{2}} \nl + 60428\,q^7 -
201792\,q^{\frac{15}{2}}
 + 707426\,q^8 - 2550208\,q^{\frac{17}{2}}
\nl + 9101288\,q^9 - 32568832\,q^{\frac{19}{2}} +
116831861\,q^{10} + \mathcal{O}(q^{\frac{21}{2}}) \;.\nn
\label{zso10A} \eea
 that can be reorganized in the form
\bea {\cal Z}_{SO(10,2)}(q)\eq (q^1)^2+(10 q^2-q^3)^2
+(-16 q^{5/2} + 54 q^3 - 10 q^4)^2\nn\\
&& + (45 q^3 - 144 q^{\frac{7}{2}} + 210 q^4 + 16 q^{\frac{9}{2}}
- 54 q^5)^2+\ldots \;, \eea It is not difficult to recognize that
\be {\cal Z}^{({\cal N}=4)}_{SO(10,2)}(q) = \sum_{\ell}
V_{\ell}^L(q) \times V_{\ell}^R(q) \ee where $q$ keeps track of
the dimensions assigned via $\Delta_0= J +\nu$ after lifting $SO(9)$
to $SO(10)$. We thus find perfect agreement with the string spectrum at the 
HS symmetry point, thus
lending support to our assumptions and extrapolations. The origin
of the $SO(10,2)$ spectrum symmetry calls for deeper understanding
possibly in connection with Bars's two-time formulations of the
superstring \cite{Bars:2004dg}.

In order to set the stage for interactions that lead to HS
symmetry breaking, one has to decompose the spectrum of
single-trace operators in free ${\cal N}=4$ SYM at large $N$ or,
equivalently, of type IIB superstring on $AdS_5\times S^5$
extrapolated to the point of HS symmetry  into HS multiplets
\cite{Beisert:2004di}. To this end, following
\cite{Vasiliev:2004qz, Sezgin:2001zs} we need to recall some basic
properties of the infinite dimensional HS (super)algebra
$hs(2,2|4)$, that extends the ${\cal N}=4$ superconformal
algebra $psu(2,2|4)$. The latter can be realized in terms of
(super-)oscillators $\zeta_\Lambda = (y_a, \theta_A)$ with $[y_a,
\bar{y}^b] = \delta_a{}^b{}$ and $ \{\theta_A, \bar{\theta}^B\} =
\delta^B{}_A $. $y_a, \bar{y}^b$ are bosonic oscillators with
$a,b=1,...4$ a Weyl spinor index of $SO(4,2)\sim SU(2,2)$,
while $\theta_A,\bar{\theta}^B$ are fermionic oscillators with
$A,B=1,...4$ a Weyl spinor index of $SO(6)\sim SU(4)$.

Generators of $PSU(2,2|4)$ are `traceless' bilinears of
superoscillators: $J^a{}_b = \bar{y}^a y_b - \ft14 K \delta^a{}_b$
with $K = \bar{y}^a y_a$, $T^A{}_B = \bar{\theta}^A \theta_B -
\ft14 B \delta^A_B$ with $B = \bar{\theta}^A \theta_A$, ${\cal
Q}^A_a = \bar{\theta}^A y_a$, and $\bar{\cal Q}^a_A = \bar{y}^a
\theta_A$. The central element $C \equiv  K+B =
\bar{\zeta}^\Lambda \zeta_\Lambda$ generates an abelian ideal that
can be modded out \eg by consistently assigning $C=0$ to the
elementary SYM fields and their (perturbative) composites. The
hypercharge $B$ acts as an external automorphism of
$psu(2,2|4)$.

The HS extension $hs(2,2|4)$ is roughly speaking generated by
odd powers of the above generators \ie
\begin{eqnarray}
hs(2,2|4) &=& \oplus_\ell\, {\cal A}_{2\ell+1} =
\sum_{\ell=0}^\infty \Big\{ {\cal J}_{2\ell+1}= P^{\Lambda_1\ldots
\Lambda_{2\ell+1}}_{\Sigma_1\ldots \Sigma_{2\ell+1}}\,
 \bar{\zeta}^{\Sigma_1}\!\dots\bar{\zeta}^{\Sigma_{2\ell+1}}\,
 \zeta_{\Lambda_1}\!\dots \zeta_{\Lambda_{2\ell+1}}
\Big\}\;,\label{phs}
\end{eqnarray}
with elements ${\cal J}_{2\ell+1}$ in ${\cal A}_{2\ell+1}$ at
level $\ell$ parameterized by (graded) traceless rank
$(2\ell\!+\!1)$ symmetric tensors $P^{\Lambda_1\ldots
\Lambda_{2\ell+1}}_{\Sigma_1\ldots \Sigma_{2\ell+1}}$. More
precisely, one first considers the enveloping algebra of
$psu(2,2|4)$, which is an associative algebra and consists of
all powers of the generators, then restricts it to the odd part
which closes as a Lie algebra modulo the central charge $C$, and
finally quotients the ideal generated by $C$. It is easy to show
that $B$ is never generated in commutators (but $C$ is!) and thus
remains an external automorphism of $hs(2,2|4)$
\cite{Sezgin:2001zs}.

To each element of $hs(2,2|4)$ with spins $(j_L,j_R)$ is
associated an HS currents and a dual HS gauge field in the AdS
bulk with spins $(j_L+\ft12,j_R+\ft12)$. The $PSU(2,2|4)$
quantum numbers can be read off from (\ref{phs}) by expanding the
polynomials in powers of $\theta$'s up to 4, since $\theta^5=0$.
There is a single superconformal multiplet $\mult_{2\ell}$ at each
level $\ell\geq 2$. The lowest spin cases $\ell=0,1$, \ie
$\hat{\mult}_{0,2}$, are special. They differ from the content of
doubleton multiplets $\mult_{0,2}$ by spin $s=0,1/2$ states
\cite{Sezgin:2001zs}. The fundamental representation of
$HS(2,2|4)$ turns out to coincide with the singleton
$\mult_{(0,0)[0,1,0]}^{1,0}$ of $PSU(2,2|4)$. Its HWS
$|Z\rangle$ or simply $Z$ is one of the complex scalars of ${\cal
N}=4$ SYM. Any state $A$ in this representation can be found by
acting on the vacuum $Z$, or any other state $B$, with a sequence
of superconformal generators. Looking at the singleton as an irrep
of $HS(2,2|4)$ the sequence of superconformal generators
connecting $B$ to $A$ is replaced by a single HS generator
$\mathcal{J}_{A\bar B}$. This property is crucial in proving 
irreducibility of the YT-pletons with respect to the HS algebra.
Indeed the tensor product of $L\geq 1$ singletons  is generically
reducible not only under $PSU(2,2|4)$ but also under
$HS(2,2|4)$ since the HS generators ${\cal J}_{2\ell+1}\equiv
\sum_{s=1}^L\, {\cal J}^{(s)}_{2\ell+1}$, being completely
symmetric, commute with (anti)symmetrizations of the indices. The
tensor product thus decomposes into a sum of representations
characterized by Young tableaux $YT$ with $L$ boxes. To prove
irreducibility of $L$-pletons associated to a specific YT's under
$HS(2,2|4)$, it is enough to show that any state in the
$L$-pleton under consideration can be found by acting on the
relevant HWS with HS generators. This is easy for the totally
symmetric YT. More effort is needed to extend the argument to
generic YT's \cite{Beisert:2004di}.

Only a subset of YT's, those compatible with cyclicity
of $SU(N)$ traces, enters the generating
function
of single-trace operators in ${\cal N}=4$ SYM theory
\[
\Yboxdim4pt {\cal Z}(q)= \sum_{n\geq 2}\, {\cal Z}_n (q)=
\sum_{n\geq 2,d|n}\, \frac{{\cal E}(d)}{n}\,{\cal
Z}_{\yng(1)}\,(q^d)^{\frac{n}{d}} \;, \label{polyah2}
\]
of cyclic words of length $L=n$. Observe that $\Yboxdim4pt {\cal
Z}_{\yng(1)}\,(q^d)$ can be rewritten as the alternating sum over
length-$d$ YT's of hook type:
\[
\Yboxdim4pt {\cal Z}_{\yng(1)}\,(q^d) ~=~ {\cal
Z}_{\yng(5)\cdot\cdot\yng(2)}(q)~-~ {\cal
Z}_{\yng(4,1)^{\cdot\cdot\yng(2)}}(q)~+~ {\cal
Z}_{\yng(3,1,1)^{^{\cdot\cdot\yng(2)}}}(q)~-~ {\cal
Z}_{\yng(2,1,1,1)^{^{^{\cdot\cdot\yng(2)}}}}(q) ~+~ \ldots \;.
\label{hook}
\]
 Plugging this expansion into
(\ref{polyah2}), we find for the first few cases:
\begin{eqnarray}
\Yboxdim4pt {\cal Z}_2 &=& \Yboxdim4pt
{\cal Z}_{\yng(2)}\;,\nn\\[1ex]
{\cal Z}_3 &=& \Yboxdim4pt {\cal Z}_{\yng(3)}+{\cal
Z}_{\yng(1,1,1)}
\;,\nn\\[1ex]
{\cal Z}_4 &=& \Yboxdim4pt {\cal Z}_{\yng(4)}+ {\cal
Z}_{\yng(2,1,1)}+{\cal Z}_{\yng(2,2)}
\;,\nn\\[1ex]
{\cal Z}_5 &=& \Yboxdim4pt {\cal Z}_{\yng(5)}+{\cal
Z}_{\yng(3,2)}+ 2\, {\cal Z}_{\yng(3,1,1)}+ {\cal
Z}_{\yng(2,2,1)}+ {\cal Z}_{\yng(1,1,1,1,1)}\;, \qquad\mbox{etc.}
\label{hsD}
\end{eqnarray}

As anticipated, HS multiplets
associated to the tableaux $\Yboxdim6pt\yng(1,1)$,
$\Yboxdim6pt\yng(2,1)$, and two out of the three of type
$\Yboxdim6pt\yng(2,1,1)$ are projected out.
Under the superconformal group $PSU(2,2|4)$, the HS multiplet
${\cal Z}_{YT}$, associated to a given Young tableau $YT$ with $L$
boxes, decomposes into an infinite sum of multiplets. The HWS's
can be found by computing ${\cal Z}_{YT}$ and eliminating the
superconformal descendants by passing ${\cal Z}_{YT}$ through a
sort of Eratosthenes (super) sieve~\cite{Bianchi:2003wx}. In the
$PSU(2,2|4)$ notation $\mult^{\Delta_0,B}_{(j_L,j_R)[q_1,p,q_2]}$
one finds for $L=2,3$

\bea \Yboxdim4pt {\cal Z}_{\yng(2)}\eq
\sum_{n=0}^\infty\mult^{2n,0}_{(-1+n^\ast,-1+n^\ast)[0,0,0]} \;,
\eea

\bea \Yboxdim4pt {\cal Z}_{\yng(3)} \eq \sum_{n=0}^\infty
c_n\left[
\mult^{1+n,0}_{(-1+\frac{1}{2}n^\ast,-1+\frac{1}{2}n^\ast)[0,1,0]}
+ \bigbrk{\mult^{\frac{11}{2}+n,
\frac{1}{2}}_{(\frac{3}{2}+\frac{1}{2}n^\ast,1+\frac{1}{2}n^\ast)[0,0,1]}
+\mbox{h.c.}}\right] \nl +\sum_{m=0}^\infty\sum_{n=0}^\infty
c_n\left[
\mult^{4+4m+n,1}_{(1+2m+\frac{1}{2}n^\ast,\frac{1}{2}n)[0,0,0]}
+\mult^{9+4m+n,1}_{(\frac{7}{2}+2m+\frac{1}{2}n^\ast,
\frac{3}{2}+\frac{1}{2}n)[0,0,0]} +\mbox{h.c.}\right] \;, \eea
\bea \Yboxdim4pt {\cal Z}_{\yng(1,1,1)} \eq \sum_{n=0}^\infty
c_n\left[ \mult^{4+n,0}_{(\frac{1}{2}+\frac{1}{2}n^\ast,
\frac{1}{2}+\frac{1}{2}n^\ast)[0,1,0]} +\bigbrk{ \mult^{
\frac{5}{2}+n, \frac{1}{2}}_{(\frac{1}{2}n^\ast,-
\frac{1}{2}+\frac{1}{2}n^\ast)[0,0,1]} +\mbox{h.c}}\right] \nl
+\sum_{m=0}^\infty\sum_{n=0}^\infty c_n\left[
\mult^{6+4m+n,1}_{(2+2m+\frac{1}{2}n^\ast,\frac{1}{2}n)[0,0,0]} +
\mult^{7+4m+n,1}_{(\frac{5}{2}+2m+\frac{1}{2}n^\ast,
\frac{3}{2}+\frac{1}{2}n)[0,0,0]} +\mbox{h.c.}\right]\;.
 \eea

The multiplicities $c_n\equiv 1+[n/6]-\delta_{n,1~{\rm mod}~6}$
with $[m]$ the integral part of $m$, of $PSU(2,2|4)$ multiplets
inside $HS(2,2|4)$ count the number of ways one can distribute
derivatives (HS descendants) between the boxes in the tableaux.

In addition to the $\ft12$-BPS with $n=0$, the symmetric doubleton
$\Yboxdim4pt{\cal Z}_{\yng(2)}$, corresponding to the quadratic
Casimir $\delta_{ab}$, contains the multiplets of conserved HS
currents $\mult_{2n}$. The antisymmetric doubleton
$\Yboxdim4pt{\cal Z}_{\yng(1,1)}$  is ruled out by cyclicity of
the trace, cf.~\eqref{hsD}.  The `symmetric tripleton'
$\Yboxdim4pt{\cal Z}_{\yng(3)}$, corresponding to the cubic
Casimir $d_{abc}$, contains the first KK recurrences of twist 2
semishort multiplets, the semishort-semishort series $\mult_{\pm
1,n}$ starting with fermionic primaries and long-semishort
multiplets. The antisymmetric tripleton $\Yboxdim4pt{\cal
Z}_{\yng(1,1,1)}$, corresponding to the structure constants
$f_{abc}$, on the other hand contains the Goldstone multiplets
that merge with twist 2 multiplets to form long multiplets when
the HS symmetry is broken, in particular, fermionic
semishort-semishort multiplets and long-semishort multiplets
\cite{Beisert:2004di}.

The holographic formulation of {\it La Grande Bouffe} we have in
mind is of the St\"uckelberg type \cite{Bianchi:2000sm, Bianchi:2001de,
Bianchi:2001kw}. Let us illustrate it for a broken singlet vector
current. The Lagrangian describing the bulk Higgs mechanism \`a la
St\"uckelberg should be (schematically) of the form \be L = -
{1\over 4} F(V)^2 + {1\over 2} (\partial \alpha - M V)^2 \ee where
$F$ is the field-strength of the  bulk vector field $V$ dual to
the current $J$ and $\alpha$ is the bulk (pseudo)scalar dual to
the `anomaly' $A =
\partial_\mu J^\mu$. Gauge invariance under \be \delta V_m =
\partial_m \vartheta \quad , \quad \delta \alpha = M  \vartheta\ee
is manifest for constant $M$. For $M=0$, $V$ and $\alpha$
decouple. For $M\neq 0$, $V$ eats $\alpha$ and becomes massive. In
practice $M$ should depend on the dilaton and other massless
scalars. Since we want to preserve superconformal invariance, $M$
can at most acquire a constant vev and the above analysis seems
valid, at least for a vector current. Although little is known
about massless HS bosonic and fermionic fields with mixed symmetry
\cite{Vasiliev:2004qz, Sorokin:2004ie, Bouatta:2004kk}, whenever
they are part of the HS doubleton multiplet, supersymmetry should
be enough to determine their equations from the more familiar
equations for symmetric tensors. By the same token, the various
manifestations of the symmetry breaking mechanism should be
related by (extended) supersymmetry. For instance, for the ${\cal
N}=4$ Konishi multiplet the axial anomaly is part of an on-shell
anomaly supermultiplet \cite{Bianchi:2000vh, Bianchi:2001cm,
Heslop:2003xu} \be \bar{D}^A \bar{D}^B {\cal K}_{long} = g_{ym}
Tr({\cal W}_{EF}[{\cal W}^{AE} {\cal W}^{BF}]) + {g_{ym}^2\over
8\pi^2} D_E D_F Tr({\cal W}^{AE} {\cal W}^{BF}) \quad .\ee

For symmetric tensors of rank $s$ in any dimension $D$, the set of
St\"uckelberg fields that participate in the spontaneous breaking
of HS symmetry can be elegantly derived performing a formal KK
reduction of the (quadratic) HS lagrangian from $D+1$ dimensions
\cite{Bouatta:2004kk}. The {\it a priori} complex Fourier modes
$\psi_{(s-t)}^M(x)\exp(i M y)$ with $t=0,...s$ exactly account for
the correct number of d.o.f.'s $\nu_{M\neq 0} (D,s)$. Indeed it is
easy to check that $\nu_{M\neq 0} (D,s) = \nu_{M=0} (D,s) +
\nu_{M\neq 0} (D,s-1)$ and, by iteration, that $\nu_{M\neq 0}
(D,s) = \sum_{t=0}^s \nu_{M=0} (D,t)$. After reduction, \ie
integration over $y$, one can take real combinations
$\phi_{(s-t)}$ of $\psi_{(s-t)}$'s. More explicitly, from a
massless doubly traceless spin $s$ field $\Phi_{(s)}$ in $D+1$
dimensions one gets `massless' fields $\phi_{(s-t)}$ with $t=0,...
s$ satisfying certain trace conditions in $D$ dimensions. The
resulting HS field equations are invariant by construction under
gauge transformations $\delta\phi_{(s-t)} = \partial_{(1)}
\epsilon_{(s-t-1)} + t M \epsilon_{(s-t)}$ with $t=0,... s$,
resulting from $\delta \Phi_{(s)} = \partial_{(1)} {\cal
E}_{(s-1)}$ with restricted (traceless) parameters that expose the
role of the lower spin fields in {\it La Grande Bouffe}.

\section*{Acknowledgements}
It is a pleasure for me to thank Niklas Beisert, Jos\'e Francisco
Morales Morera and Henning Samtleben for a very fruitful and
stimulating collaboration. Let me take this chance to acknowledge
long lasting collaborations with Dan Freedman, Mike Green, Stefano
Kovacs, Giancarlo Rossi, Kostas Skenderis and Yassen Stanev that
have contributed significantly to shape my understanding of the
holographic correspondence. Most of what I know on HS fields I
learnt from Misha Vasiliev, Per Sundell, Ergin Sezgin, Augusto
Sagnotti, Fabio Riccioni, Tassos Petkou and Dario Francia. This
work was supported in part by I.N.F.N., by the EC programs
HPRN-CT-2000-00122, HPRN-CT-2000-00131 and HPRN-CT-2000-00148, by
the INTAS contract 99-1-590, by the MURST-COFIN contract
2001-025492 and by the NATO contract PST.CLG.978785.

\end{document}